\documentclass[%
 reprint,twocolumn,floatfix,
 amsmath,amssymb,
 aps,
 prx,
longbibliography,
superscriptaddress
]{revtex4-2}

\usepackage{graphicx}% Include figure files
\usepackage{dcolumn}% Align table columns on decimal point
\usepackage{bm}% bold math
\usepackage{color}
\usepackage{comment}
\usepackage{url}
\usepackage[normalem]{ulem}
\usepackage[version=3]{mhchem} % for reaction equations

\usepackage{amsmath}
\usepackage{amssymb}
\usepackage{lipsum}
\usepackage{url}
\usepackage{breakurl}
\usepackage[breaklinks]{hyperref}
\usepackage{comment}

\renewcommand{\epsilon}{\varepsilon}
\graphicspath{{Figs/}}
\begin{document}

\title{Capillary imbibition in lubricant coated channels}
\author{Sergi G. Leyva}
%\email{sergi.granados@ub.edu}
\affiliation{Departament de F\'{i}sica de la Mat\`{e}ria Condensada, Universitat de Barcelona, 08028 Spain}
\affiliation{Universitat de Barcelona Institute of Complex Systems (UBICS), Universitat de Barcelona, 08028 Spain} 
\author{Ignacio Pagonabarraga}
\affiliation{Departament de F\'{i}sica de la Mat\`{e}ria Condensada, Universitat de Barcelona, 08028 Spain}
\affiliation{Universitat de Barcelona Institute of Complex Systems (UBICS), Universitat de Barcelona, 08028 Spain}
\author{Aurora Hern\'andez-Machado}
\affiliation{Departament de F\'{i}sica de la Mat\`{e}ria Condensada, Universitat de Barcelona, 08028 Spain}
\affiliation{Institut de Nanoci\`{e}ncia i Nanotecnologia, Universitat de Barcelona, 08028 Spain}
\author{Rodrigo Ledesma-Aguilar}
\affiliation{School of Engineering, University of Edinburgh, The King's Buildings, Mayfield Road, Edinburgh EH9 3JL, UK}
\date{\today}
\begin{abstract}

{
Capillary imbibition underpins many processes of fundamental and applied relevance in fluid mechanics. A limitation to the flow is the coupling to the confining solid, which induces friction forces. Our work proposes a general theoretical framework for the modeling of the transport of liquids in lubricant impregnated surfaces. We show that for sufficiently small lubricant viscosity, dissipation entirely occurs in the lubricant layer, resulting in a linear growth of the advancing front. As a result, an external force gives rise to an exponential front growth. This new capacity to control multiphase flows sets new experimental challenges that can be determinant for micro and nanofluidic devices.

}

\end{abstract}

\maketitle

Spontaneous imbibition, also known as capillary filling, occurs when one fluid  displaces a second one from a solid porous medium due to its preferential affinity to wet the internal surfaces of the solid. 
Applications can be found in nanofluidics~\cite{Rauscher2008}, where elastocapillary forces support the self-assembly of arrays of carbon nanotubes \cite{nanotubes};
%
%\textcolor{red}{which can result in cellular patterns %\cite{nanotubes2}. 
%}
%
biophysics, where capillary forces are known to influence protein folding \cite{bico_2018,folding,protein2};  
and medical devices, where lateral flows, an example of capillary driven flows, are widely used to detect the presence of a target substance, and set the basis for antigen detection \cite{Li2012,lateral_flow}.

Classical imbibition corresponds to a viscous fluid displacing a gas in a uniform porous medium, where the front position, $l(t)$, follows the “slowing-down” growth of Washburn’s law, $l(t) \propto t^\alpha$, with $\alpha=1/2$~\cite{quereburn1921}. Understanding and controlling the exponent $\alpha$, is therefore of both fundamental and practical interest. Pradas et al.~\cite{prades2006,prades2008} and Queralt et al.~\cite{queralt2011} showed that the exponent can be lowered to $\alpha<1/2$ by introducing disorder in the channel topography. On the other hand, Primkulov et al.~\cite{juanes2020} reported a larger exponent $\alpha=1$, but lower imbibition speed, by capping the front with a slug of a viscous oil.

%The interaction of the advancing and receding fluids with the solid constitutes one of the main limiting factors on the front advance.
%
%Upon contact with the porous medium, the invading fluid is driven in by the surface tension of the meniscus, which is opposed by the friction imparted by the solid surface.
%
%For a liquid displacing a gas in a uniform porous medium,
%
%the viscous friction of the invading liquid grows with increasing penetration length, 
%
%leading at long times to the ``slowing-down'' dynamics of Washburn's law, $l(t) \propto t^{1/2}$~\cite{washburn1921}, where $l$ is the position of the advancing front and $t$ is time. 

What happens if the solid walls of a porous medium are replaced by a liquid surface? 
Experimental realisations of lubricant impregnated surfaces, like Liquid-Infused Porous Surfaces (SLIPS) or Lubricant-Impregnated Surfaces (LIS)~\cite{aizenberg2011,smith2013droplet,SLIPS_1,LIS_1,LIS_2} have gained much attention in the recent years. 
These materials have outstanding properties for droplet manipulation due to their low friction, resistance to extreme conditions, and self-healing properties, as well as their ability to induce drag reduction in contact with a single liquid phase 
\cite{slips_fish,chiara_2022,slips_drag}. 
Here we address the fundamental question of how spontaneously invades a porous medium coated with a lubricant.

We show that the lubricant viscosity plays a determinant role to trigger a qualitative change in the dissipation mechanism where the liquid front advances at a constant rate, instead of slowing down, as would occur if the fluids were in direct contact with the solid. 
We also identify  the high sensitivity of the liquid front to external perturbations, which opens an avenue to new modes of liquid manipulation in the microscale. 

\begin{figure*}[]
\centering
%\subfloat{
\begin{flushleft}
%\hspace{0.11\textwidth}{\large{(a)}} \\
\begin{center}
 \includegraphics[width=\textwidth,keepaspectratio]{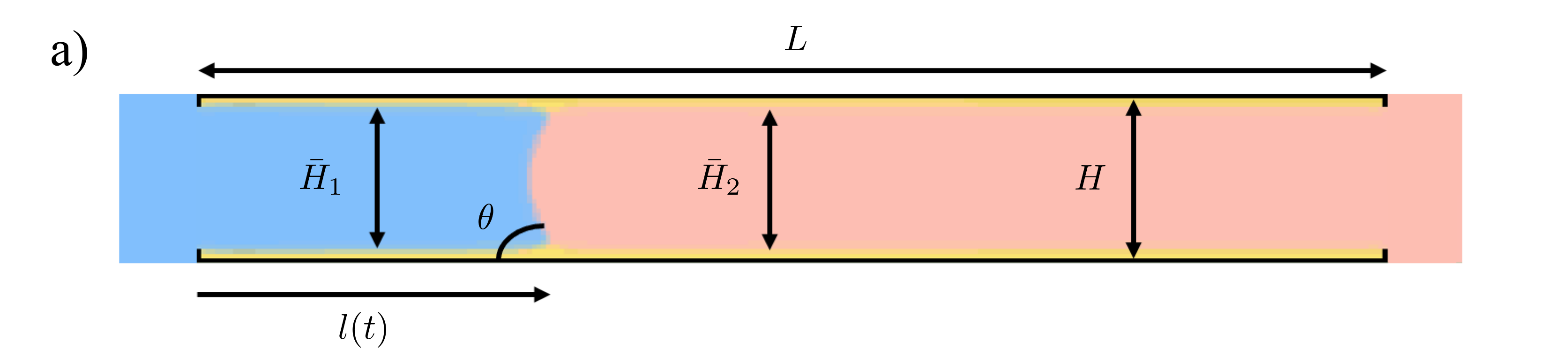}
\end{center}
\end{flushleft}
%}
%\begin{flushleft}
% \hspace{0.11\textwidth}{\large{(b)}} \hspace{0.45\textwidth}{\large{(c)}}\\
% \end{flushleft}
%\subfloat{
%\hspace{0.08\textwidth}
\begin{center}
\includegraphics[width=\columnwidth]{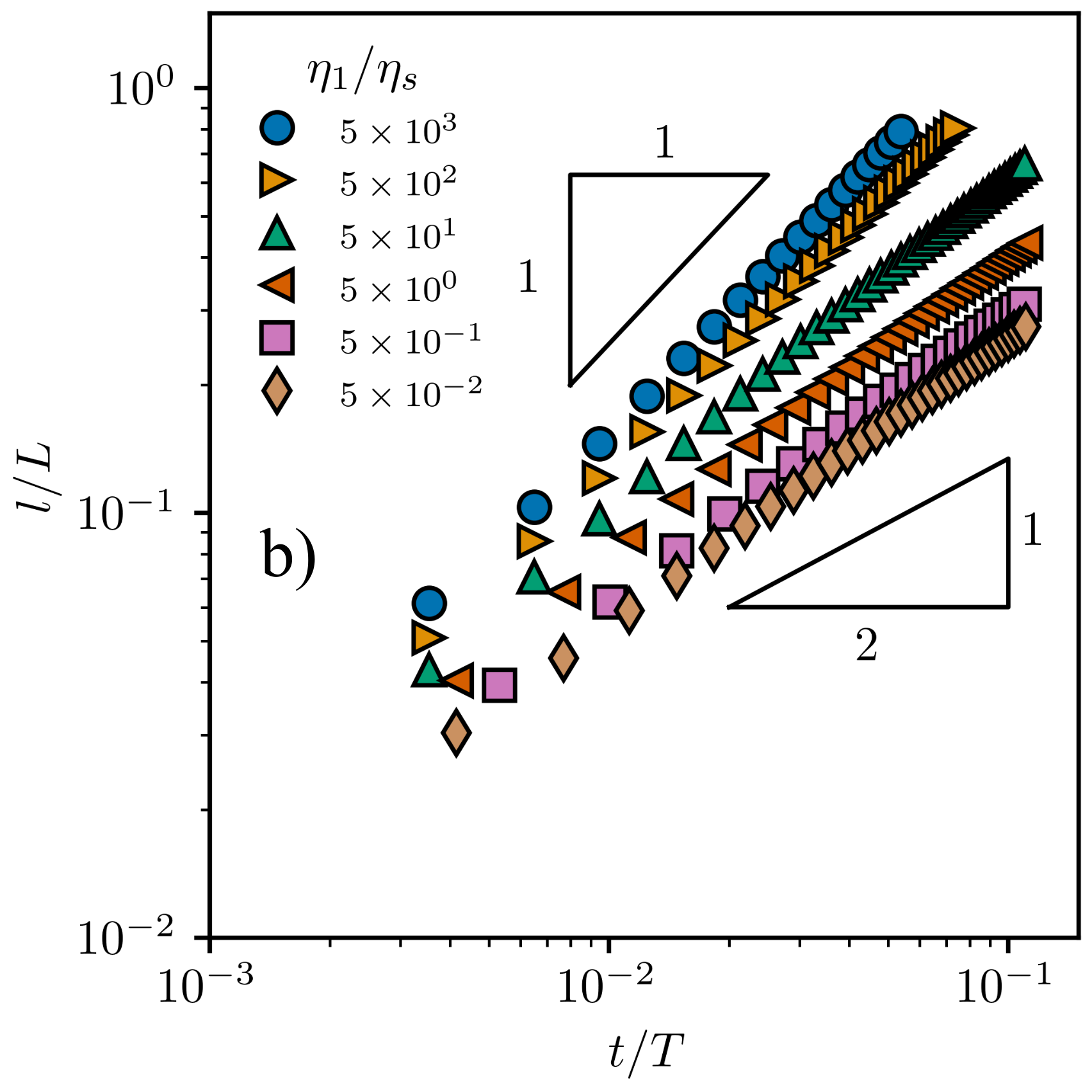}
%}
%\subfloat{
\includegraphics[width=\columnwidth]{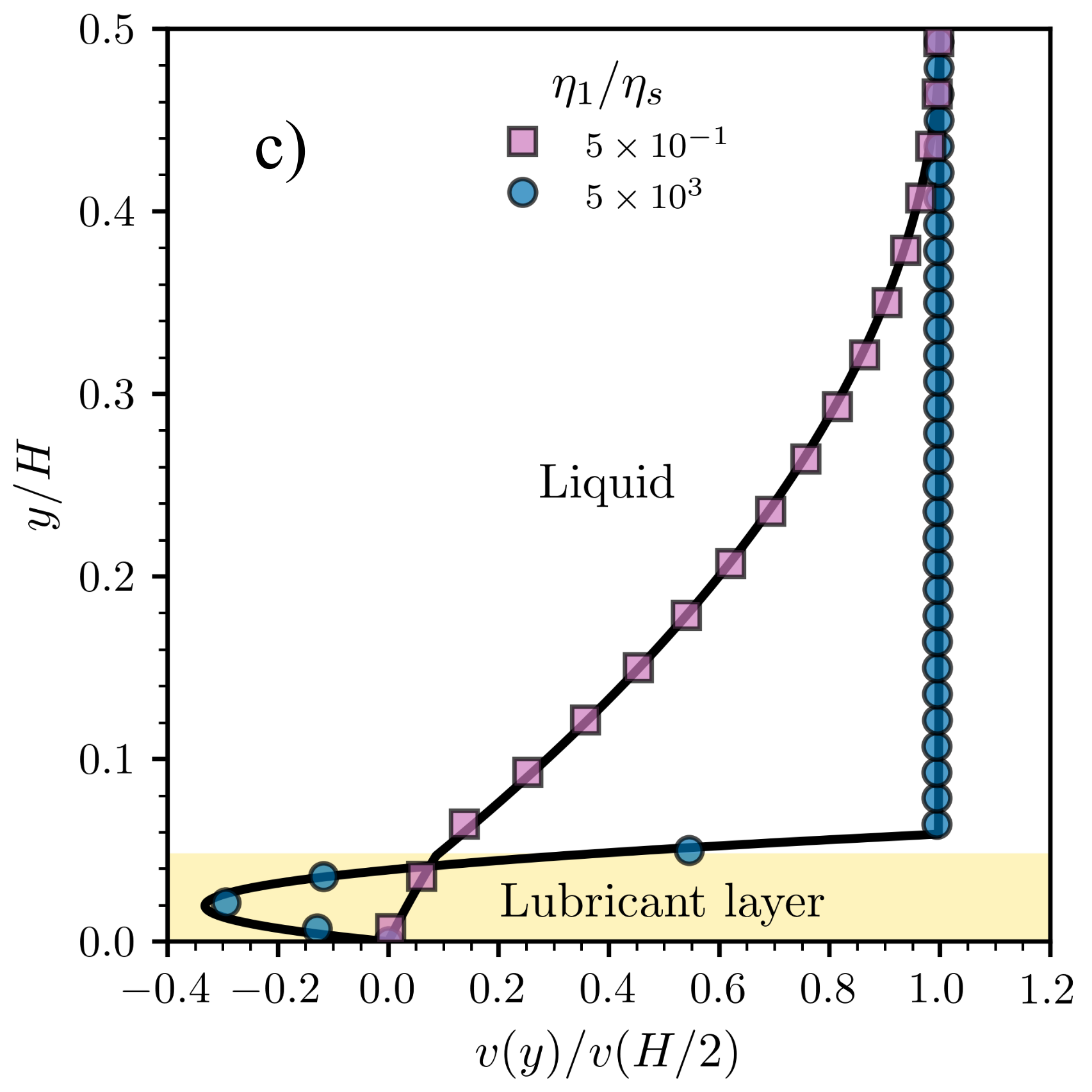}
\end{center}
%}

\caption{ Spontaneous imbibition in a lubricant coated channel. (a) Lattice-Boltzmann simulation snapshot. 
The liquid on the left, of width $\bar H_1$, preferentially wets the surface of a thin lubricant layer of width $\bar h_1$, and displaces a resident fluid in a channel of width $H=\bar H_1+2h_1$, and length $L$. 
A meniscus, of apparent angle $\theta$ and position $l(t)$, advances within the channel. 
(b) Effect of the viscosity of the lubricant, $\eta_s$, on the imbibtion curves at fixed viscosity contrast between the displaced and displacing liquids, $\eta_2/\eta_1=1\times10^{-2}$. 
(c) Velocity profile in the displacing fluid and in the lubricant layer, $v(y)$, for  
$\eta_1/\eta_s=5\times{10}^{-1}$ (squares) and $\eta_1/\eta_s=5\times{10}^{3}$ (circles).
The solid lines shows the theoretical prediction (see text). 
The velocity is made dimensionless by the velocity at the center of the channel, $v(H/2)$, while the $y$ coordinate is normalised by the channel width, $H$. 
\label{figure1}}

%{\the\textwidth}
\end{figure*}

To elucidate the front dynamics, we have carried out 2D lattice-Boltzmann (LB) numerical simulations of the imbibition into a solid channel coated with a liquid lubricant layer. We couple the LB method to a ternary free-energy model of three immiscible fluids, which we solve by using the Cahn-Hilliard equation. The ternary free-energy model~\cite{halim_ternary}, allows to independently choose the surface tensions of the liquids, and thus the Neumann angles at their intersection as well as the wettability of the solid.
In our simulations, Fig.~\ref{figure1}a, two reservoirs hold liquids of equal density but different viscosities,  $\eta_1$ and $\eta_2$. 
The reservoirs are connected by a solid channel of length $L$ and width $H$, whose internal surfaces are coated by a thin film of a third liquid of viscosity $\eta_s$ (the lubricant).
The lubricant is kept in place by two small pillars located at the edges of the channel. This geometry mimics SLIPS, where the lubricant is locked into the surface by roughness~\cite{glaco}. 
Further details on the simulation methodology and choice of parameters  are provided in the supplementary information~\cite{EPAPS}. 

As shown in the simulation snapshot of Fig.~\ref{figure1}a, an appropriate choice of the surface tensions leads to the spontaneous imbibition of liquid 1 into the channel, displacing liquid 2. 
The displacing and displaced liquids form an advancing meniscus with a well-defined apparent angle relative to the solid, $\theta$, 
suggesting a driving capillary force $F_c \propto \gamma \cos \theta$, where $\gamma$ is the surface tension of the interface between liquids 1 and 2.
Despite this similarity, the meniscus does not touch the solid, but moves on top of the lubricant layer.

%%%%%%%%%%%%%%%%%%%%%%%%%%%%%%%%
% Fig. 1
%%%%%%%%%%%%%%%%%%%%%%%%%%%%%%%
%\begin{figure*}[ht]

We focus on the familiar case of a viscous liquid displacing a much less viscous fluid, $\eta_2/\eta_1 = 10^{-2}$, and analyse the motion of the meniscus at different lubricant viscosities, $\eta_s$. 
Fig.~\ref{figure1}b shows the corresponding $l$ vs $t$ curves, where the penetration length is normalised by the length of the tube, and time is normalised using the filling time predicted by Washburn's law, $T={3\eta_1{L}^2}/{H\gamma\cos{\theta}}$~\cite{ledesma_2022}.
For very large lubricant viscosity, 
the front advances following the scaling of Washburn's law, $l(t)\sim t^{1/2}$, indicating that the viscous force, $F_v$, increases with increasing $l$.
Decreasing the lubricant viscosity leads to an unexpected result: The front grows linearly, $l(t)\sim t$, thus suggesting that $F_v$ is independent of $l$.
In addition, the filling time is significantly shorter than that predicted by Washburn's law.
We shall show that these effects are not a transient due to inertia or dynamic-angle effects~\cite{JOOS1990,blake_2003,Qur1997InertialC,ledesma_2022}. 
They correspond to a new long-time regime entirely dominated by the viscous dissipation in the lubricant film.

Fig.~\ref{figure1}c shows profiles of the tangential velocity of the displacing phase and the lubricant far upstream of the meniscus.
The expected parabolic flow profile of a forced fluid~\cite{bruus2008} is approached for large $\eta_s$. 
The velocity in the lubricant layer becomes vanishingly small, which effectively behaves like a solid. 
In contrast, for small $\eta_s$, the flow profile resembles a plug flow in the displacing and displaced phases, while there is strong variation of the velocity in the lubricant layer, where shear stresses are sustained.

%
%%%%%%%%%%%%%%%%%%%%%%%%%%%%%%%%
% Fig. 2
%%%%%%%%%%%%%%%%%%%%%%%%%%%%%%%
%\begin{figure}[t!]
\begin{figure}[]
%\begin{flushleft}
% {(a)} 
%\end{flushleft}
\begin{center}
\includegraphics[width=\columnwidth]{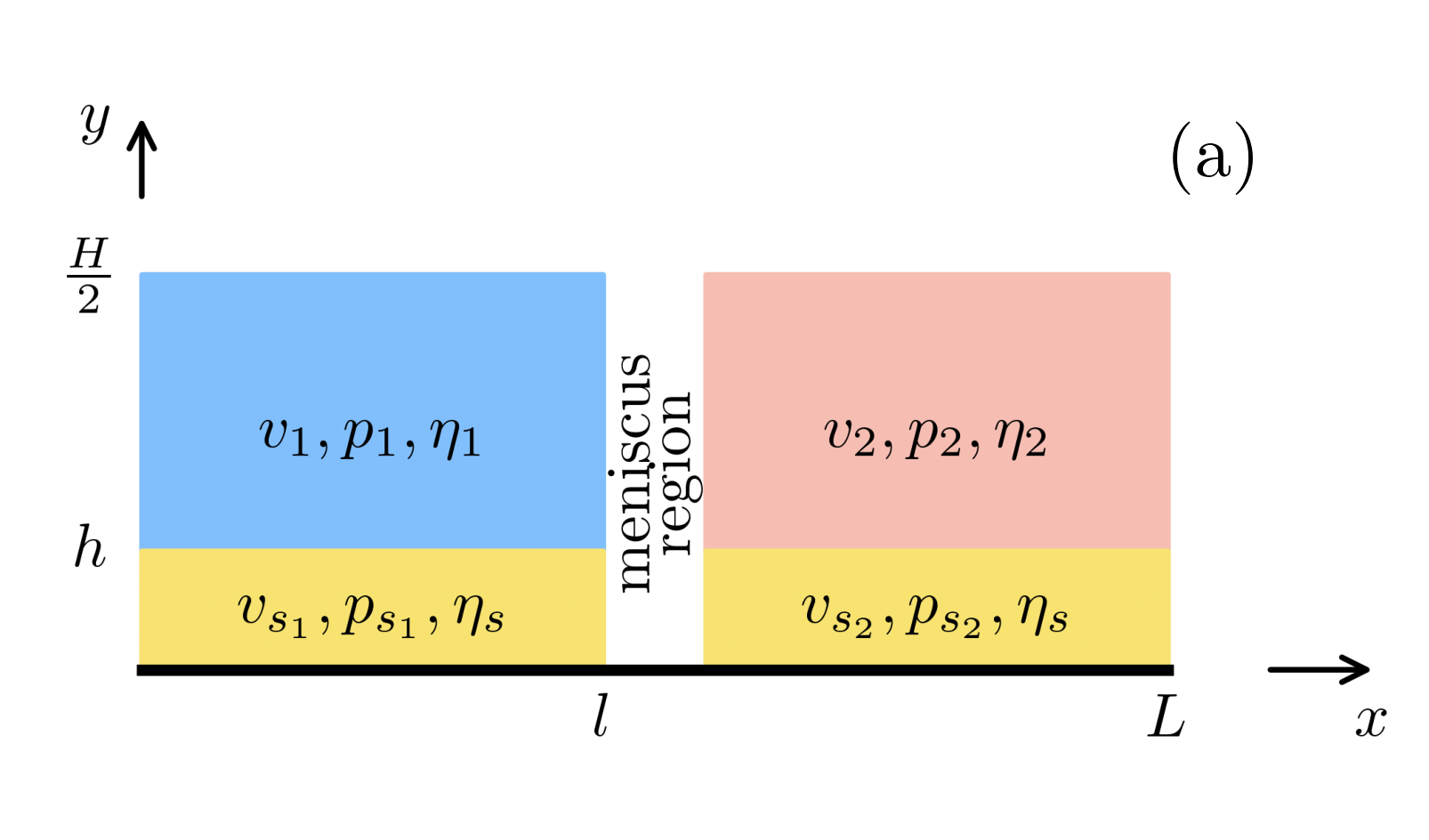}

\includegraphics[width=\columnwidth]{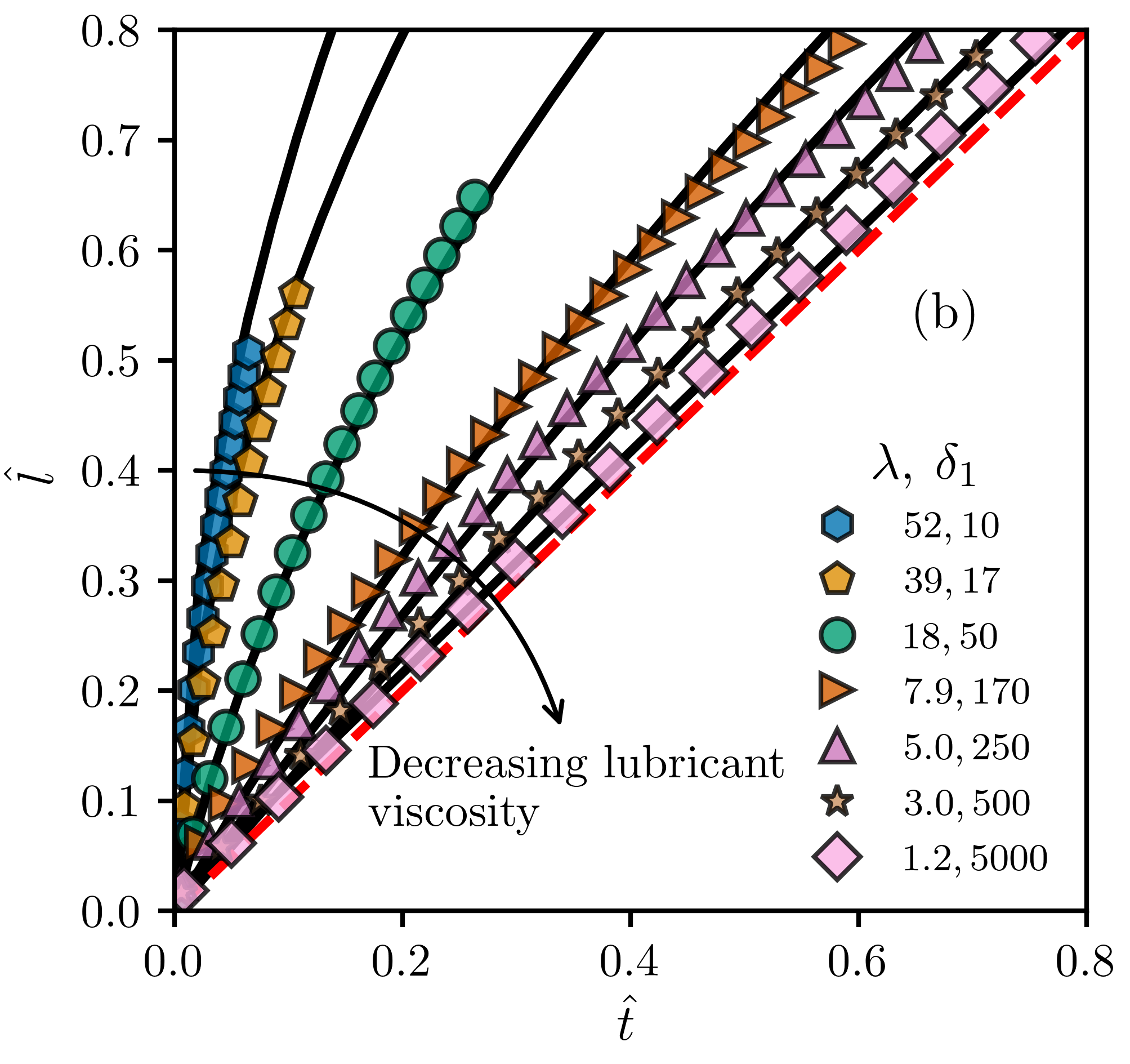}
\end{center}
	\caption{ (a) Schematics of the lubricant coated channel. The rectangular sections represent the bulk of the displacing and displaced liquids, and of the lubricant layer, which are separated by the meniscus region. (b) Theoretical imbibition curves (solid lines) vs simulation results (symbols) for different values of the parameter $\lambda$. For $\lambda\rightarrow 1$, corresponding to small lubricant viscosity, the imbibition curves approach the asymptotic limit $\hat{l}=\hat{t}$ (red dashed line). The values of $\Sigma$ with decreasing lubricant viscosity correspond to $0.03, 0.01, 0.2, 0.9,1.7, 3.2$ and $24$, respectively.   
	\label{figure2}}
	%\the\columnwidth
\end{figure}
%%%%%%%%%%%%%%%%%%%%%%%%%%%%%%% 
%

Building on these observations, we propose a simplified model of the flow in each fluid phase where
we neglect the dynamics close to the meniscus and describe the flow profile in the four ``bulk'' flow regions depicted in Fig.~\ref{figure2}a.
The tangential velocity profiles in the displacing and displaced phases, $v_1(y)$ and $v_2(y)$, and in the adjacent regions of the lubricant layer, $v_{s1}(y)$ and $v_{s2}(y)$ are obtained from lubrication theory, assuming   
that the pressure profiles in each region, namely $p_1$, $p_2$, $p_{s1}$ and $p_{s2}$, only vary along the longitudinal coordinate $x$. 
Accordingly, the flow profiles obey the Stokes equations.
Four out of the eight integration constants are found by imposing continuity of
the velocity and tangential stress at the interface with the lubricant layer.
The remaining constants are determined by fixing the average velocities of the fluids and the lubricant, $\frac{1}{H/2-h}\int_h^{H/2}v_{i}dy=u$ and
$\frac{1}{h}\int_0^hv_{s_i}dy=u_s$ \cite{EPAPS}.

In general, $u$ and $u_s$ are independent free parameters; however,  for the imbibition geometry the lubricant layer responds to the capillary driving force that acts on the meniscus.
Therefore, we expect that if no external forces in the lubricant are present, the average velocity of the lubricant obeys $u_s=\alpha  u$, with $0\leq \alpha \leq 1$. 
This relationship allows us to eliminate the pressure gradient terms from the Stokes equations, 
and instead characterize the flow through  $u$ and $\alpha$. 
Figure~\ref{figure1}c shows the excellent agreement of the theoretical prediction with the simulation velocity profiles, where $u_s$ and $u$ are fixed to the measured liquid flow. 
The theoretical results show that a vanishing average lubricant velocity $u_s$ leads to a negative derivative of the velocity profile close to the solid, indicating a recirculating flow in the lubricant layer, as reported in SLIPS/LIS simulations~\cite{chiara_2022} (see analytical solution in the supplementary material \cite{EPAPS} ). 

The model can be used to determine 
the viscous friction force (per unit length) exerted by the lubricant layer on the moving fluids,
{$F_v=2l\tau_1+2(L-l)\tau_2$}, where $\tau_i\equiv \:\eta_s dv_{s_i}(h)/dy$ is the shear stress. 
From the velocity profiles we obtain \cite{EPAPS}
\begin{equation}
 \label{eq:all_frictions_equation}
 F_v=\frac{4(2-3\alpha)u\eta_s}{h}\left(\frac{l}{1+\frac{2\bar{H}}{3\delta_1{h}}}+\frac{L-l}{1+\frac{2\bar{H}}{3\delta_2h}}\right).
 \end{equation} 
Here, {$\bar H_i\equiv H-2h= \bar H$} is the width of either liquid and 
 $\delta_i\equiv \eta_i/\eta_s, i=1,2$ is the viscosity ratio between the displacing/displaced ($i=1$/$i=2$) liquid and the lubricant.
Letting $\delta_i h/\bar H\rightarrow{0}$,  $\eta_2/\eta_1\rightarrow 0$ and setting $\alpha=0$, 
this expression reduces to the classical result of the viscous force acting on a single liquid in contact with a solid channel  
i.e.,  $F_v ={12 \eta_1 u l}/{H}$.
On the contrary, the limit of small lubricant viscosity is achieved by letting $\delta_i h/\bar H\rightarrow\infty$, where the friction force reduces to  $F_v={4(2-3\alpha)\eta_s u L}/{h}.$ 
In this regime the viscous force is dominated by the lubricant layer, despite  being the {\it less} viscous phase, and 
the force does not depend on the position of the front; 
rather, its magnitude scales with the entire length of the channel, $L$.  
Comparing the energy dissipation rate in the bulk of the displacing and displaced liquids $\dot{E}_b=\int_0^l\int_h^{H/2}\eta_1|{\nabla}v_{1}|^2\:dydx+\int_l^L\int_h^{H/2}\eta_2|{\nabla}v_{2}|^2\:dydx,$ 
to that of the lubricant $\dot{E}_s=\int _0^l\int_0^h\eta_s|{\nabla}v_{s_1}|^2\:dydx+\int _l^L\int_0^h\eta_s|{\nabla}v_{s_2}|^2\:dydx,$ for $\delta_ih/\bar H\rightarrow\infty$, we find ${\dot{E}_b}/{\dot{E}_s}\rightarrow 0$ \cite{EPAPS}.
Therefore, in this limit the energy dissipation occurs in the lubricant, and not in the bulk of the displacing and displaced phases. 

The  imbibition growth law, $\dot{l}(t)$, is derived from  the force balance, $F_c=F_v+F_m$,  between the capillary ($F_c$), viscous ($F_v$) and contact-line friction ($F_m$) forces per unit length. 
The simulations show that  the contact angle settles to a constant value, $\theta\approx \theta_e$ 
after a short transient.
Accordingly, $F_c=2\bar{H}\gamma\cos\theta_e/H$. 
 The unbalanced interfacial stress close to the triple line is given by $\gamma\cos\theta(u)$. Since the contact angle is finite, for small velocities, one can expand this term $\gamma\cos\theta(u)\simeq\gamma\cos\theta(0)+ku$, resulting in $F_m=ku$, where $k$ is a friction coefficient \cite{ledesma_2022}. Independently, it has been reported that such a linear scaling holds in the limit $\eta_s/\eta_1\rightarrow 0$  \cite{quere2017}.

Using the dimensionless variables $\hat l \equiv l/L$, $\hat t \equiv t\gamma\cos\theta\bar{H}/LH[4\eta_sL(2-3\alpha)/h+k]$ and $\hat u\equiv d \hat l /d\hat t$, 
we can integrate the equation of the front motion and obtain 
\begin{equation}
\label{eq:diff_equation}
\frac{\hat{l}^2}{2}\frac{(\lambda-1)}{1+\frac{2\bar{H}}{3h\delta_1}}
+\hat{l}\left(\frac{1}{1+\frac{2\bar{H}}{3h\delta_1}}+\lambda{\Sigma}\right)
=\lambda\left(1+{\Sigma}\right)\hat t,
\end{equation}
where
\begin{equation}
\lambda\equiv\frac{\eta_1}{\eta_2}\frac{3\delta_2h+2\bar{H}}{3\delta_1h+2\bar{H}}\quad\textrm{~and~}\quad{\Sigma}\equiv \frac{kh}{4\eta_s{L}(2-3\alpha)}.
\label{eq:parameters}
\end{equation}
The parameter $\lambda$ contains the relative effect of the viscosities of the three fluids together with the fraction of the channel occupied by the lubricant, and 
$\Sigma$ quantifies the strength of the friction of the meniscus relative to the lubricant layer.

As shown in Fig.~\ref{figure2}a, Eq.~\ref{eq:diff_equation} agrees well with the simulations, with the  contact-line friction coefficient, $k$, used as the only fitting parameter \footnote{The value of $k=34$ simulation units, is computed once, which indicates it has a weak dependence on the lubricant properties and is controlled by the viscosities in the displacing and displaced fluids}.
 The classical, diffusive-like growth regime of Wahsburn's Law is recovered by imposing $\Sigma=0$ and letting $\delta_i h/\bar H\rightarrow 0$. 
This eliminates the effect of the meniscus and reduces $\lambda$ to the familiar viscosity contrast, i.e.,  $\lambda \rightarrow \eta_1/\eta_2$. Then, taking $\lambda\gg1$ gives $\hat l \rightarrow \sqrt{4\bar{H}\eta_s \hat t/3h\eta_1}$. 

On the other hand, for $\delta_i h/\bar H\gg1$ and $\lambda\rightarrow 1$  equation~(\ref{eq:diff_equation}) yields the linear growth law $\hat{l}\rightarrow \hat{t}$. After recovering dimensions we find
\begin{equation}
l(t)=\frac{\bar{H}h\gamma\cos\theta}{H[4\eta_s{L}(2-3\alpha)+kh]}t.
\end{equation}
Remarkably,  the velocity of the front depends only on the viscosity and thickness of the lubricant layer and on the channel length.
The linear growth regime identified in this letter occurs for $\delta_ih/\bar H\gg1$, where energy dissipation occurs primarily in the lubricant layer.
For  intermediate regimes the asymptotic growth of the front  will conform to Washburn's law.
A cross-over length $l_c$ can be estimated by comparing the magnitudes of quadratic and linear contributions in Eq.~\ref{eq:diff_equation}.
We obtain $l_c \sim2L(1+\lambda\Sigma[1+2\bar{H}/3h\delta_1])/(\lambda-1)$, which implies $l_c > L$ if $\lambda<3/(1-2\Sigma[1+2\bar{H}/3h\delta_1])$. 
If the  friction associated to the contact line is negligible compared to the lubricant dissipation $\Sigma=0$,  a crossover length $l_c>L$ requires $\lambda<3$. 
However, increasing the contact line friction increases the crossover length, and
 when ${\Sigma}>1/2$ Washburn's law will never be observed as the simulation results show in Fig.~\ref{figure2}b.

\begin{figure}[]
\includegraphics[width=\columnwidth]{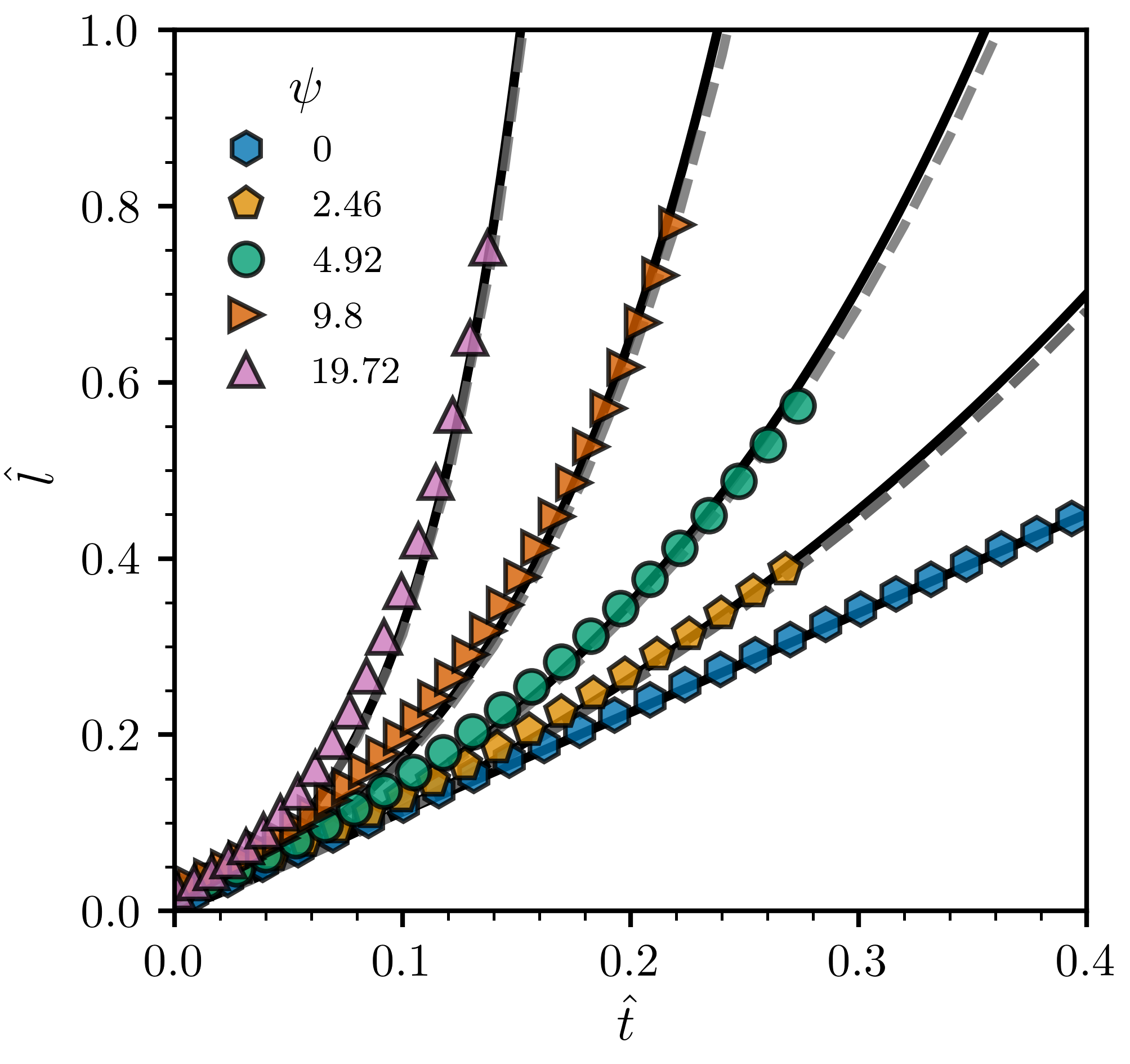}
\caption{Forced imbibition dynamics in a lubricant-coated channel. 
External forcing leads to an exponential growth of the position of the advancing front. 
The strength of the external force is quantified the forcing coefficient $\psi$. The simulation parameters correspond to $\Sigma=24$, $\lambda=1.2$ and $\delta_1=5000$.
The symbols correspond to simulations, solid lines correspond to the numerical solution of the differencial equation including the forcing term, and the dashed grey lines to the approximation in Eq.~\ref{eq:exponential_grav}.\label{figure3}}
\end{figure}

Here we have focused on the case of spontaneous imbibition.
However, a low-viscosity lubricant layer has a strong impact in the sensitivity of the front to perturbations, and  leads to a significant modification of the front dynamics when the fluids are subject to  external forces.
Let us consider a uniform external force acting on the displacing liquid, $F_e=f\bar{H}l$, for small $\eta_s$. 
Fig.~\ref{figure3} shows a speed-up of the front as it advances in the channel, in sharp contrast to the classical result of forced imbibition, where the motion of the front is linear.
This effect is also captured by the theoretical model upon adding an external force, which leads to the solid curves shown in Fig.~\ref{figure3} after numerical integration. 
An approximate expression of the growth law can be obtained in the regime $\lambda\rightarrow 1$ and $\delta_1\rightarrow\infty$, which gives 
\begin{equation}
\label{eq:exponential_grav}
\hat{l}=\frac{e^{\psi\hat{t}}-1}{\psi}=\frac{e^{t/t_i}-1}{\psi},
\end{equation}
where $\psi=LHf/2\gamma\cos\theta$ is the forcing coefficient. 
Equation~(\ref{eq:exponential_grav}) predicts an exponential invasion of the channel in agreement with the simulation results (dashed curved in Fig.~\ref{figure3}), with a characteristic time scale $t_i=\eta_sL/Hhf$ determined by the competition between viscous forces in the lubricant layer and the external forcing. 

Our work can be used to design experimental setups that optimize imbibition in SLIPS/LIS channels. For example, using liquids with viscosities $\eta_1=1700$ mPas, $\eta_2=17$ mPas and $\eta_s=10$ mPas \cite{slips_review}, and a typical lubricant thickness of 1~$\mu$m in a channel of $H\simeq{20}~\mu$m would result in $\lambda\simeq{8}$, thus making the linear regime reported here accessible in experiments.
%
%%%%%%%%%%%%%%%%%%%%%%%%%%%%%%%%
% Fig. 3
%%%%%%%%%%%%%%%%%%%%%%%%%%%%%%%
%\begin{figure}[b!]
%Experimental realisation of the latter, coupled to its capacity to transport multiphase flows at a constant rate,  sets a new experimental challenge that can be determinant for nano and microfluidic devices and LIS/SLIPS aplications.
The analytical framework, validated with simulations, provides a starting point to characterise further effects that might be relevant in spontaneous imbibition processes in SLIPS and LIS, such as the precise role of the dissipation in the ridge, or the effect of the varying width of the lubricant in the channel. 

Altogether, the ideas reported in this work will help rationalise the effect of a lubricant layer in naturally-occurring situations as well as inspire solutions to technological challenges. 
For example, in pitcher plants, which inspired SLIPS originally~\cite{aizenberg2011}, the textured surface that supports the lubricant layer has corrugations which form semi-open channels, and such structures could sustain the capillary flows reported in this paper~\cite{pitch_plant}. 
Binary liquid capillary bridges, which spontaneously move in confinement and can therefore be used for droplet transport applications,  have been reported to spontaneously leave a thin film of liquid adhered to a channel wall~\cite{bico2002self}.
In antifouling applications~\cite{antifouling_1,antifouling3,antifouling_4,antifouling_5,quere2017}, some studies suggest that bacteria can accumulate in the lubricant layer, limiting its medical applications \cite{antifouling_2}. Controlling spontaneous flows could be a key solution to enhance LIS properties and make it a suitable material for medical applications \cite{detaching}.
 %
%Here we have studied the case of Newtonian liquids, however the dynamics of viscoelastic liquids in bilayer configurations~\cite{bilayer} could introduce the additional effect of an intrinsic timescale to the dynamics.}

%\Acknowledgements 
I. P. acknowledges support from Ministerio de Ciencia, Innovaci\'on y
Universidades (Grant No. PID2021-126570NB-100  AEI/FEDER-EU), from
Generalitat de Catalunya  under Program Icrea Acad\`emia and project 2021SGR-673DURSI (Grant
No. 2017 SGR 884).
A.H.-M. acknowledges support from Ministerio de Ciencia, Innovaci\'on y
Universidades (Grant No. PID2019-106063GB-100).

\bibliography{Bibliography}

\end{document}

\title{Supplementary Material for: \\ Capillary imbibition in lubricant coated channels}
\author{Sergi G. Leyva}
\email{sergi.granados@ub.edu}
%\thanks{Both authors equally contributed to this work}
\affiliation{Departament de F\'{i}sica de la Mat\`{e}ria Condensada, Universitat de Barcelona, 08028 Spain}
\affiliation{Universitat de Barcelona Institute of Complex Systems (UBICS), Universitat de Barcelona, 08028 Spain} 
\author{Ignacio Pagonabarraga}
\affiliation{Departament de F\'{i}sica de la Mat\`{e}ria Condensada, Universitat de Barcelona, 08028 Spain}
\affiliation{Universitat de Barcelona Institute of Complex Systems (UBICS), Universitat de Barcelona, 08028 Spain}
\affiliation{Centre Europ\'een de Calcul Atomique et Mol\'eculaire (CECAM), \'Ecole Polytechnique F\'ed\'erale de Lausanne (EPFL) , 1015 Lausanne, Switzerland}
\author{Aurora Hern\'andez-Machado}
\affiliation{Departament de F\'{i}sica de la Mat\`{e}ria Condensada, Universitat de Barcelona, 08028 Spain}
\affiliation{Institut de Nanoci\`{e}ncia i Nanotecnologia, Universitat de Barcelona, 08028 Spain}
\author{Rodrigo Ledesma-Aguilar}
\affiliation{School of Engineering, University of Edinburgh, The King's Buildings, Mayfield Road, Edinburgh EH9 3JL, UK}
\date{\today}
\maketitle

\section{Details of the simulations}

The simulations are based in the Lattice-Boltzmann (LB) method, where the discretised distribution function $f_i$ defined at each node of a regular mesh, and each velocity $i$ of a set of velocities follows 

\begin{equation}
 f_i(\boldsymbol{r}+\boldsymbol{c}_i\Delta{t};t+\Delta{t})=f_i(\boldsymbol{r};t)+\sum_j{L_{ij}(f_i(\boldsymbol{r};t)-f_i^{eq}(\boldsymbol{r};t))} \, \, , 
\end{equation}
where $\boldsymbol{c}_i$ is the discrete velocity basis and $\Delta{t}$ is the discrete time step. $L_{ij}$ is the collision operator that provides the way by which the distribution function evolves towards its equilibrium value. Given a velocity set the discrete distribution function provides a series of moments that relate to physical quantities of the model $\rho(\boldsymbol{r};t)=\sum_if_i(\boldsymbol{r};t), \rho{u}_{\alpha}(\boldsymbol{r};t)=\sum_if_i(\boldsymbol{r};t)c_{i\alpha}, \Pi_{\alpha\beta}(\boldsymbol{r};t)=\sum_if_i(\boldsymbol{r};t)c_{i\alpha}c_{i\beta}$. The summation takes place over the discrete set of velocities that corresponds to a certain basis, and the number of each basis depends on the basis election. The number of discrete velocities is called $N_{vel}$. In our method we will generally use three dimensions and 19 discrete velocities, which is widely known as the D3Q19 set. Further details on the collision operator and implementation can be found in the documentation of the implementation we used, Ludwig, which is an open access code [1]. 
The three liquid phases are implemented by means of a ternary free energy [2]

%\begin{equation}
%\label{eq:free_energy}
\begin{equation}
\begin{split}
F=\int_{\Omega}\frac{\kappa_1}{2}C_1^2(1-C_1)^2+\frac{\kappa_2}{2}C_2^2(1-C_2)^2+ \\ \frac{\kappa_s}{2}C_3^2(1-C_s)^2+\frac{\kappa'_1}{2}(\nabla{C_1})^2+\frac{\kappa'_2}{2}(\nabla{C_2})^2+\frac{\kappa'_s}{2}(\nabla{C_s})^2,
\end{split}
\end{equation}

where the subindex $s$ refers to the fluid phase we use as the lubricant.

In our current implementation, the order parameters representing each phase are evolved following the Cahn-Hilliard equation
\begin{equation}
\partial_t{C_i}=\partial_\alpha(C_iu_\alpha-M\partial_\alpha\mu_i)
\end{equation}
which involves calculation of the advection of the order parameter, and the gradient of the chemical potential for each phase $i$. We set $\kappa_i=\beta\kappa'_i$ to impose that all interfaces have the same width. 
The solid-liquid surface tension can be tuned by means of introducing a boundary condition of the gradient of the order parameters normal to solid boundaries, through the wetting parameters $h_1$ and $h_2$. 
With this model, one can tune the equilibrium contact angles between the different fluids $\theta_1$, $\theta_2$ and $\theta_s$, through $\kappa_i$ and $\alpha$,   and each combination of fluids and solid through the wetting parameters $h_1$ and $h_2$, $\theta_{12}$, $\theta_{1s}$, $\theta_{s2}$. 
All the details of this calculations relating the contact angles and the free energy parameters can be found in Ref.~[2]. 

We briefly introduce the set of parameters used for the simulations. The total length and width of the channel are $L=500$ and $H=70$, respectively. We add two large reservoirs, of displacing (fluid 1, left) and displaced (fluid 2, right) phases, to allow for the displacing phase to invade the whole length of the capillary (from left to right). 
The upper and lower lubricant layers (fluid $s$) are initialised with a uniform width of $h=4$ fluid nodes, and the displacing-displaced interface is set at an initial front position $l(0)=0.02L$. The viscosity contrast is $\eta_2=\eta_1/100$, and we choose a viscosity $\eta_1=5$, which in our simulations ensures stability in a range of $\eta_s$ of orders of magnitude $\eta_s\in(0.001,50)$.
We choose the surface tension of the lubricant liquid with the other two to lad to contact angle close to 180 degrees, to minimize the effect of the curvature of the interface when the three liquids meet. More specifically, we choose $\kappa_1=0.002$, and $\kappa_2=\kappa_3=0.007$, which leads to an equilibrium angles between fluids of $77.8^\circ$, $141.1^\circ$ and $141.1^\circ$ for fluids 1, s, and 2, respectively. 
For stabilising the lubricant layer, we choose complete wetting of the lubricant respect the displacing and displaced fluids by means of
satisfying the relations $S_{1s}=\cos\theta_{1s}-\gamma_{1s}>1$ and $S_{s2}=\cos\theta_{s2}-\gamma_{s2}<-1$. This is achieved by a set of wetting parameters $h_1=1.75\cdot{10^-4}$ and $h_1=1.75\cdot{10^-3}$. Furthermore, this wetting parameters lead to an equilibrium contact angle $\theta_{12}=51^{\circ}$ that is hydrophilic in order to enhance spontaneous imbibition of the pore. When the lubricant is introduced, the interaction of the front between the lubricant and the solid leads to a measured contact equilibrium angle of $\theta=49^\circ$, which is the angle that sets the pressure drop.

\section{Analytical solution}

Our theoretical model consists in solving the Stokes equation
\begin{equation}
\eta_i\frac{\partial^2v_{i}}{\partial{y}^2}=\frac{dp_i}{dx}, \qquad i=1,2,
\label{eq:Stokes}
\end{equation} 
and
\begin{equation}
\eta_s\frac{\partial^2v_{s_i}}{\partial{y}^2}=\frac{dp_{s_i}}{dx}, \qquad i=1,2,
\label{eq:StokesLub}
\end{equation}
for the displacing and displaced phases, 1 and 2 respectively, and the corresponding lubricant velocity profile. The boundary conditions, already introduced in the paper, consist on imposing the continuity of the velocity and tangential stress at the interface with the lubricant layer, i.e., $v_{i}(h)=v_{s_i}(h)$ and $\eta_i dv_{i}(h)/dy=\eta_sdv_{s_i}(h)/dy$. 
and fixing the average velocities of the fluids and the lubricant, $\frac{1}{H/2-h}\int_h^{H/2}v_{i}dy=u$ and
$\frac{1}{h}\int_0^hv_{s_i}dy=u_s$. With these boundary conditions, we obtain the velocity profiles
\begin{equation}
    v_{s_i}(y)=a_i y + c_i y^2 \qquad {\rm for} \qquad 0 \leq  y \leq h,
    \label{eq:vlubricant}
\end{equation}
and
\begin{equation}
    v_{i}(y)=b_i+d_i(y-H/2)^2    \qquad  {\rm for} \qquad h \leq y \leq  H/2.
    \label{eq:vcentral}
\end{equation}

where the coefficients can be determined analitically
\begin{equation*}
a=\frac{3u_s}{h}+\frac{3(3u_s-2u)\delta}{3h\delta+2\bar{H}}
\end{equation*} 

\begin{equation*}
b=\frac{6u\delta{h}+3\bar{H}(2u-u_s)}{2(3e\delta+2\bar{H})}
\end{equation*}

\begin{equation*}
c=\frac{9u\delta{h}-3u_s(6h\delta+\bar{H})}{h^2(3e\delta+2\bar{H})}
\end{equation*}

\begin{equation*}
d=\frac{6(2u-3u_s)\delta}{\bar{H}(3h\delta)+2\bar{H}}
\end{equation*}

From Eq.~6, it becomes clear that an average lubricant velocity $u_s=0$ results in recirculation of fluid, since $a=-6u\delta/(3h\delta+2\bar{H})<0$, thus resulting in a negative derivative of the flow profile at y=0. Even for small average lubricant velocities, some recirculation is expected. It can be shown that $a$ becomes zero when $u_s=uh\delta/(3h\delta+\bar{H})$, taking a positive value. For lower $u_s$, we expect some lubricant going towards the opposite direction of the front close to the solid. 

For calculating the dissipation rate between the lubricant and the bulk fluids,  we note that $\dot{E}_b=\int_0^l\int_h^{H/2}\eta_1|{\nabla}v_{1}|^2\:dydx+\int_l^L\int_h^{H/2}\eta_2|{\nabla}v_{2}|^2\:dydx,$
to that of the lubricant,  
$\dot{E}_s=\int _0^l\int_0^h\eta_s|{\nabla}v_{s_1}|^2\:dydx+\int _l^L\int_0^h\eta_s|{\nabla}v_{s_2}|^2\:dydx.$ 

Using this expressions and the analytical solution leads to 

\begin{equation}
\frac{\dot{E}_b}{\dot{E}_s}=\frac{2\bar{H}}{3h}\frac{\left(\frac{3{h}\delta_1+2\bar{H}}{3{h}\delta_2+2\bar{H}}\right)^2+1}{\delta_2^2\left(\frac{3{h}\delta_1+2\bar{H}}{3h\delta_2+2\bar{H}}\right)^2+\delta_1^2}.
\end{equation}.

For adimensionalising curves in Fig.~2b, we use the equilibrium angle $\theta=49^\circ$, and fit $k$ to the curve of smallest lubricant viscosity, which has a value of ${k}=34$ simulation units. Then we use this same value for all the curves. For each simulation, we measure the average slip width in the channel and calculate the $\Sigma$ variable to obtain the solutions from our model.

The force balance, $F_v+F_m=F_c+F_e$ leads to the equation of motion
\begin{equation}
\label{eq:diff_equation_full}
\frac{d \hat{l}}{d \hat t}\left[\frac{\hat{l}(\lambda-1)+1}{1+\frac{2\bar{H}}{3h\delta_1}}+\lambda{\Sigma}\right]=\left(1+{\Sigma}+\hat{l}\psi\right)\lambda,
\end{equation}
where $\psi$ contains the external forcing, as defined in the paper. If $\psi$ is set to 0, we obtain Eq.~2 of the manuscript. If $\psi$ is not zero, an analytical solution can be obtained in the case $\lambda=1$, which corresponds to Eq.~5 in the article. In Fig.~3, we plot the numerical solution to Eq.~9 here (solid line), and compare it to the approximate solution Eq.~5 of the article (dashed lines) to check its validity. 

%\section{Friction force at the meniscus}

%The unbalanced interfacial stress close to the triple line is given by $\gamma\cos\theta(u)$, where $\theta(u)$ is the dynamic apparent contact angle. 
%Because the contact angle is always finite, one can expand this expression as $\gamma\cos\theta(u)\simeq\gamma\cos\theta(0)+ku$. Thus, provided that the meniscus speed is small, as the spontaneous flows considered in this work, we expect the friction force to scale linearly with the velocity.

%$\gamma\cos\theta(u)\simeq\gamma\cos\theta(0)+ku$.
%The unbalanced interfacial stress close to the triple line is given by $\gamma\cos\theta(u)$

\section{Comparing simulations and theoretical results}

Here we report the values of the parameters we used for reproducing the curves in Fig. 3b). The initial lubricant width corresponds to a width $h_0=4$ and relaxes to an average constant value $h$, which we measure in the simulations and always remains close to $h_0$. The steady lubricant width is tipically a smaller than $h_0$, since some lubricant is accumulated in the meniscous due to the surface tensions and the resulting forces acting in the Neumann triangle. Since the channel width is H=70, this results in a $\bar{H}=H-2h\simeq62$ LB units. In the simulations, the average velocity of the lubricant, characterised by $\alpha$, is negligible compared to that of fluids 1 and 2. For example, for the largest lubricant viscosity $\eta_s$, we measure the largest $\alpha\simeq{0.06}$. In this case, the influence of this small flux turns out to be negligible and the curves fit using $\alpha=0$. As the lubricant viscosity decreases, for $\eta_s<1$, the lubricant flux vanishes $\alpha=0$. In the imbibition curves,  In Fig. 3b), the values of $\delta_2$ correspond, from large to small lubricant viscosities to 0.1, 0.83, 0.5, 1.7, 2.5, 5 and 50 respectively. With these parameters we can obtain the values of $\lambda$. In our simulations, the meniscous friction is characterised by a parameter k=34, which remains constant as $\eta_s$ is varied, in our simulation range, for $\delta_2$ as small as 0.1. This allows to obtain the $\Sigma$'s reported in the caption of Fig 3.